\documentclass[aps,prb,reprint,longbibliography]{revtex4-2}
\usepackage{amsmath}
\usepackage{amssymb}
\usepackage{graphicx}
\usepackage{color}
\usepackage{natbib}
\usepackage[colorlinks=true,
urlcolor=blue,
linkcolor=blue,
citecolor=blue]{hyperref}
\usepackage{url}

\newcommand{\abs}[1]{\lvert #1 \rvert}
\graphicspath{{../figures/}}

\newcommand{\bgl}{Bogoliubov }
\usepackage[mathlines]{lineno}

\begin{document}
	\title{Dynamical quantum phase transitions through the lens of mode 
		dynamics}
	\author{Akash Mitra}
	\author{Shashi C. L. Srivastava}
	\affiliation{Variable Energy Cyclotron Centre, 1/AF Bidhannagar, Kolkata 
		700064, India}
	\affiliation{Homi Bhabha National Institute, Training School Complex, 
		Anushaktinagar, Mumbai - 400094, India}
	\begin{abstract}
		We study the mode dynamics of a generic quadratic fermionic Hamiltonian 
		under a sudden quench protocol in momentum space. Modes with zero 
		energy at any given time, $t$, are referred to as dynamical critical 
		modes. Among all zero-energy modes, spin-flip symmetry is restored in 
		the eigenvector corresponding to selected zero-energy modes. This 
		symmetry restoration is used to define the dynamical quantum phase 
		transition (DQPT). This shows that the occurrence of these 
		dynamical critical modes is necessary but not sufficient for a 
		DQPT. We show that the conditions on the quench protocol and time 
		for such dynamical symmetry restoration are the same as the divergence 
		of the rate function and integer jump in the dynamical 
		topological order parameter, which have 
		been the traditional identifiers of a DQPT. This perspective also 
		naturally explains when one or both of DQPT and 
		ground-state quantum phase transitions will 	occur.
	\end{abstract}
	\maketitle

	\noindent \textbf{\emph{Introduction.--}}
	Phase transitions driven by non-thermal control parameters in quantum 
	many-body systems, where the ground state undergoes a non-analytic change, 
	are referred to as quantum phase transitions (QPTs). At the critical point 
	of a QPT, the system loses its characteristic time-scale, resulting in 
	scale 
	invariance, which leads to universal behavior 
	\cite{Vojta_2003,Sachdev_2011}. Extending the idea of studying physics due 
	to non-analyticity in certain quantities to out-of-equilibrium dynamics has 
	been taken up under the name of dynamical quantum phase transition (DQPT). 
	In this scenario, the Loschmidt amplitude, defined as the overlap between 
	the initial ground state and the time-evolved state, plays a role analogous 
	to the equilibrium partition function \cite{Heyl_2018}. In close analogy 
	with the Lee-Yang zeros \cite{LeeYang_1952}, DQPTs are characterized by the 
	zeros of the Loschmidt amplitude \cite{Heyl_2013}. These zeros give rise to 
	non-analytic behavior in the rate function, which plays the role of free 
	energy and is defined as the logarithm of the return probability, i.e., the 
	modulus squared of the Loschmidt amplitude. The search for a dynamical 
	order parameter lead to the quantity known as the dynamical topological 
	order parameter 
	(DTOP), which is now frequently used for characterizing DQPT 
	\cite{Budich_2016}. In particular, the DTOP remains constant at times when 
	the time-evolved state is non-orthogonal to the initial state, and it 
	undergoes an integer jump precisely at the critical times when a DQPT occurs
	\cite{Budich_2016,dutta_2017,Utso_2017,Yang_2019}.
	Since the introduction of DQPTs in the transverse field Ising model 
	\cite{Heyl_2013}, they have been extensively studied, particularly in spin 
	models 
	\cite{Vajna_2014,Heyl_2014,Heyl_2015,Halimeh_2017,Homrighausen_2017,Zauner_Stauber_2017,unkovi_2018,Lang_2018,Lang_2018c,Jafari_2019,Nicola_2021},
	topological models 
	\cite{Vajna_2015,Budich_2016,Sedlmayr_2018,Sadrzadeh_2021,Yu_2021}, Floquet 
	systems \cite{Sharma_2014,Yang_2019,Zhou_2021}, higher-dimensional systems 
	\cite{Schmitt_2015}, non-integrable models 
	\cite{Karrasch_2013,Hubig_2019,cheraghi_2021}, non-Hermitian systems 
	\cite{Zhou_2018,mondal_2022,mondal_2023,Jing_2024}, and have been extended 
	to mixed states \cite{bhattacharya_2017,Budich_2017,lang2018dynamical}. 
	Moreover, DQPTs have been experimentally realized in various platforms, 
	including trapped ions \cite{Jurcevic_2017}, cold atom systems 
	\cite{Flschner_2017, Muniz_2020}, and superconducting qubits 
	\cite{Guo_2019}.
	
	The dominant contribution to the partition function, written in the 
	canonical ensemble, comes from the ground state of the system in the 
	zero-temperature limit. Any non-analyticity in the behavior of the ground 
	state at quantum critical points results in non-analytic behavior of the 
	free energy. Therefore, the language of thermal phase transitions is still 
	useful for quantum phase transitions as well. In translationally invariant 
	systems, the critical points of the Hamiltonian are often identified by the 
	vanishing of the energy spectrum in momentum space for specific momentum 
	modes \cite{Bermudez_2010,Vodola_2014,Maity_2020,Ares_2015}. While the 
	connection between the vanishing of momentum-mode energy and the 
	non-analytic behavior of the free energy is straightforward for the ground 
	state, this connection is less clear for out-of-equilibrium states. To 
	investigate this relationship, we calculate the dynamical mode energies 
	(DMEs) of the time-evolved state following a sudden quench protocol in 
	this paper. In analogy with the ground-state QPT, we refer 
	to the vanishing of the DMEs as the emergence of dynamical critical 
	modes. We critically examine these modes, which are defined as the 
	instantaneous eigenstates of the dynamically evolved pre-quench 
	Hamiltonian. The symmetry property of these instantaneous eigenvectors 
	with respect to a $\mathbb{Z}_2$ symmetry corresponding to the spin-flip 
	operation becomes the defining property for DQPT. From this 
	perspective, we explain the earlier results of DQPTs 
	accompanied by a ground-state QPT, as well as cases where one occurs 
	without the other. To capture this symmetry restoration in the eigenstates 
	of dynamically critical modes, we introduce a quantity $\mathcal{R}(t) $ in 
	Eq. \ref{eq:Rate} and show its equivalence with the rate function 
	regardless of the quench protocol. Thus, providing an alternate view of 
	rate function in terms of spin-flip symmetry restoration of the eigenstates 
	of critical dynamical modes.
	
	\noindent \textbf{\emph{Dynamical mode energy.--}}
	The translationally invariant quadratic fermionic Hamiltonian defined on a one 
	dimensional lattice of length $N$ is given by,
	\begin{equation}\label{eq:gen_form_Ham}
		\begin{aligned}
			H = J\sum_{p,q} \Big[ 
			& t_{|p-q|}\,  c_p^\dagger c_q + t_{|p-q|}^* \, c_q^\dagger c_p \\
			& + \frac{1}{2} \left( \Delta_{|p-q|} \, c_p^\dagger c_q^\dagger + 
			\Delta_{|p-q|}^* \, c_q c_p \right)
			\Big],
		\end{aligned}
	\end{equation}
	where $c_p^\dagger$ ($c_q$) are fermionic creation (annihilation) operator. 
	The coefficients $t_{|p-q|}$ and $\Delta_{|p-q|}$ represent the hopping 
	and pairing amplitudes, respectively, and $*$ denote their complex 
	conjugates. 
	The parameter $J$ determines the overall energy 
	scale. 
	Due to anti-commuting nature of fermionic operators, 
	$\Delta$ must be an anti-symmetric matrix. 
	Utilizing the translation invariance, the Hamiltonian in momentum space can 
	be written as,
	\begin{equation}\label{eq:gen_momspace}
		H=\frac{J}{2}\sum _{n=0}^{N-1} \begin{bmatrix}
			c_{k_n}^\dagger& c_{-k_n}
		\end{bmatrix}\begin{bmatrix}
			\epsilon_{k_n} & \Delta_{k_n}^*\\\Delta_{k_n} & -\epsilon_{k_n}^*
		\end{bmatrix}\begin{bmatrix}
			c_{k_n}\\ c_{-k_n}^\dagger
		\end{bmatrix},
	\end{equation}
	where $\epsilon_{k_n}=\sum_r t_r e^{i k_n r}$ and $\Delta_{k_n}=\sum_r 
	\Delta_r e^{i k_n r}$ and $r$ denotes the distance between two sites. Owing 
	to the antiperiodic boundary condition, the lattice momenta $k_n$ are 
	quantized as $k_n=\frac{2 \pi}{N} (n+1/2)$. Connecting the $c-$fermions 
	with 
	\bgl ($\gamma$) fermions using, 
	\begin{align}\label{eq:transform_Bogolyubov}
		\begin{bmatrix}
			c_{k_n} \\
			c_{-k_n}^\dagger
		\end{bmatrix}= \begin{bmatrix}\cos \theta_{k_n} & i\sin \theta_{k_n} \\
			i \sin \theta_{k_n} & \cos \theta_{k_n}
		\end{bmatrix}\begin{bmatrix}
			\gamma_{k_n} \\
			\gamma_{-k_n}^\dagger
		\end{bmatrix},
	\end{align}
	with $\tan(2\theta_{k_n})=i\Delta_{k_n}^*/\epsilon_{k_n}$ ,
	the Hamiltonian 
	in Eq.~\ref{eq:gen_form_Ham} can be brought to the following diagonal 
	form: 
	\begin{equation}\label{eq:gs_gen_ham}
		H=J\sum_n \lambda_{k_n} \left(\gamma_{k_n}^\dagger \gamma_{k_n} 
		-\frac{1}{2}\right),
	\end{equation}
	with individual mode energy, 
	$\lambda_{k_n}=\sqrt{|\epsilon_{k_n}|^2+|\Delta_{k_n}|^2}$. Without any 
	loss of generality, we hereafter  set the overall energy scale $J=2$. The 
	ground state of the above Hamiltonian is \bgl vacuum 
	$|\psi(0)\rangle$, where $\gamma_{k_n} |\psi(0)\rangle=0$ $\forall \,k_n$. 
	The quantum critical points of the ground state correspond to the vanishing 
	of momentum mode energy which results in gapless excitations.
	
	Under a sudden quench protocol, where one or more parameters of the 
	Hamiltonian are suddenly changed, the time evolved state at each time $t$, 
	can still be mode resolved. The expectation value of the pre-quench 
	Hamiltonian can be written as the sum of energy contributions from the 
	different modes. These energies are referred to as dynamical mode energies 
	(DMEs),	$\Tilde{\lambda}_{k_n}(t)$, such that
	\begin{equation}
		\langle \psi(t)|H_0|\psi(t)\rangle =\sum_n  
		\Tilde{\lambda}_{k_n}(t)=\Tilde{\lambda}(t),
	\end{equation} 
	The spectrum of the pre-quench Hamiltonian, denoted as 	
	$\lambda_{k_n}^{(0)}$ 
	encodes ground state properties of $H_0$, including quantum critical points 
	which can be identified by mode softening, i.e., $\lambda_{k_n}^{(0)} \to 0$ at 
	critical momentum mode $k_n=k_c$. Borrowing the definition from ground state 
	analogue, zeroes of DMEs will signify \emph{dynamical criticality}. 	
	Under a unitary time evolution, the time-evolved state $|\psi(t)\rangle$ is 
	given by $|\psi(t)\rangle=e^{-iHt} |\psi(0)\rangle$, where $H$ is the 
	post-quench Hamiltonian and $|\psi(0)\rangle$ is the initial state. We take 
	ground state of the pre-quench Hamiltonian as $|\psi(0)\rangle$. As a result, $ 
	\Tilde{\lambda}(t)$ can be expressed as
	\begin{equation}\label{eq:htilde}
		\Tilde{\lambda}(t)= \langle \psi(0)| e^{i H t} H_0 e^{-i H 
			t}|\psi(0)\rangle= \langle \psi(0)| \Tilde{H}(t)|\psi(0)\rangle,
	\end{equation}
	where $\Tilde{H}(t)=e^{i H t} H_0 e^{-i H t}$. To compute 
	$\Tilde{\lambda}(t)$, we express the post-quench \bgl operators in terms of 
	pre-quench \bgl operators as, 
	\begin{align}
		\begin{bmatrix}
			\gamma_{k_n}\\
			\gamma^\dagger_{-k_n}
		\end{bmatrix}
		= \begin{bmatrix}
			\cos \delta\theta_{k_n} & -i \sin \delta \theta_{k_n}\\
			-i \sin \delta \theta_{k_n} & \cos \delta \theta_{k_n}
		\end{bmatrix}
		\begin{bmatrix}
			\gamma_{k_n}^{(0)}\\
			\gamma^{\dagger (0)}_{-k_n}
		\end{bmatrix},
	\end{align}
	where $\delta \theta_{k_n}=\theta_{k_n}-\theta_{k_n}^{(0)}$ is the 
	difference between post-quench and pre-quench Bogoliubov angles. Using this 
	relation, $\tilde{H}(t)$ can be expressed in terms of the pre-quench \bgl 
	fermions as:
	\begin{equation}\label{eq:H_tilde}
		\Tilde{H}(t)
		= -\sum_{n=0}^{N-1}\frac{\lambda_{k_n}^{(0)}}{2} 
		\begin{bmatrix}
			\gamma_{k_n}^{\dagger(0)} & \gamma_{-k_n}^{(0)}
		\end{bmatrix}
		\bigl(\vec{d}_{k_n}(t)\cdot\vec{\sigma}\bigr)
		\begin{bmatrix}
			\gamma_{k_n}^{(0)} \\[4pt] 
			\gamma_{-k_n}^{\dagger(0)}
		\end{bmatrix},
	\end{equation}
	where $\vec{\sigma}=(\sigma_x.\sigma_y,\sigma_z)$ are the Pauli matrices in the 
	pseudo-spin space. Here, the components of the time-dependent pseudo-spin 
	vector, $\vec{d}_{k_n}(t)=\left(d^x_{k_n}(t),d^y_{k_n}(t),d^z_{k_n}(t)\right)$ 
	are obtained as
	\begin{equation}
		\begin{aligned}
			d^x_{k_n}(t)&=-\sin 2 \delta \theta_{k_n} \sin \left(2\lambda_{k_n} 
			t\right) 
			\\ d^y_{k_n}(t)&=\sin 2\delta \theta_{k_n} \cos 2 \delta \theta_{k_n} 
			\left(1- \cos (2 \lambda_{k_n} t)\right)\\
			d^z_{k_n}(t)&= \cos \left(2 \lambda_{k_n} t\right)\left(\cos^2 \left(2 
			\delta 
			\theta_{k_n}\right)-1\right)-\cos^2 \left(2 \delta \theta_{k_n}\right).
		\end{aligned}
	\end{equation}
	We recall that since $|\psi(0)\rangle$ is ground state of the pre-quench 
	Hamiltonian, the lower diagonal term of 
	$-\frac{\lambda^{(0)}_{k_n}}{2}\vec{d}_{k_n}(t)\cdot\vec{\sigma}$ will give the 
	$\tilde{\lambda}_{k_n}(t)$ i.e., $\tilde{\lambda}_{k_n}(t) = 
	\frac{\lambda^{(0)}_{k_n}}{2} d^z_{k_n}(t)$. The dynamical criticality 
	therefore means $d^z_{k_n}(t)=0$,  which implies that, at a given time and for 
	a given momentum mode, the corresponding pseudo-spin lies perpendicular to the 
	$z$-axis ( i.e. in $xy-$plane).  Any state in this plane can therefore  be 
	written as  $\frac{1}{\sqrt{2}}[|\uparrow\rangle + 
	e^{i\Phi_{k_n}(t)}|\downarrow\rangle]$, where $\Phi_{k_n}(t)$ is the azimuthal 
	angle when Pauli vector $\vec{d}_{k_n}(t)$ is represented in spherical-polar 
	coordinates. This azimuthal angle can be expressed as
	\begin{equation}\label{eq:azimuth}
		\begin{aligned}
			\Phi_{k_n}(t)&=\tan^{-1}\frac{d^y_{k_n}(t)}{d^x_{k_n}(t)} = 
			\tan^{-1}\left[- \cos (2 \delta \theta_{k_n}) \tan \left(\lambda_{k_n} 
			t\right)\right].
		\end{aligned}
	\end{equation}
	For a zero polar angle of the Pauli vector, $\vec{d}_{k_n}(t)\cdot\vec{\sigma}$ 
	is proportional to $\sigma_z$ and the ground state is denoted by 
	$|\uparrow\rangle$. At the other extreme of the polar angle, since the sign of 
	$\vec{d}_{k_n}(t)\cdot\vec{\sigma}$ is reversed, the ground state will be 
	$|\downarrow\rangle$.
	
	Now, let's define a symmetry of flipping these eigenvectors from 
	$|\uparrow\rangle$ to $|\downarrow\rangle$ and vice versa, then initial state 
	(i.e. pre-quenched  state) is in a symmetry broken phase. Let's assume that at 
	a certain time $t=t_c$, we obtain a symmetry-restored phase i.e. 
	$\frac{1}{\sqrt{2}}[|\uparrow\rangle +|\downarrow\rangle]$, for a particular 
	$k_n=k_c$. This occurs when  
	$\Phi_{k_c}(t_c)=0$ i.e. $d^y_{k_c}(t_c) = 0, d^x_{k_c}(t_c) \neq 0$, along 
	with $d^z_{k_c}(t_c) = 0$. We define this dynamical transition as the 
	\emph{dynamical quantum phase transition}. These conditions, in terms of the 
	quench parameter and time, imply:
	\begin{equation}
		\begin{aligned}\label{eq:dqpt_cond}
			\delta \theta_{k_c} = \left(m_1 + \frac{1}{2}\right)\frac{\pi}{2}, ~
			t_c = \left(m_2 + \frac{1}{2}\right)\frac{\pi}{2 \lambda_{k_c} }, m_i \in 
			\mathbb{N}^0.
		\end{aligned}
	\end{equation}
	Note that a vanishing DME becomes a necessary condition for a DQPT. However, as 
	we will see later in detail, one can choose 
	time and modes in such a way that 
	$d^z_{k_n}=0$, but Eq. \ref{eq:dqpt_cond} is not satisfied. These 
	solutions will give vanishing DME but not a DQPT.  It is important to 
	explicitly state that 
	these conditions (Eq.~\ref{eq:dqpt_cond}) correspond exactly to the divergence of 
	the
	rate function and integer jumps in the DTOP \cite{supp} 
	\nocite{heyl_2019, 
		Pancharatnam_1956},  both 
	of which are well-studied quantifiers theoretically 
	\cite{Heyl_2013,Vajna_2014,Heyl_2014,Heyl_2015,Halimeh_2017,Homrighausen_2017,
		Zauner_Stauber_2017,unkovi_2018,Lang_2018,Budich_2016,dutta_2017,Utso_2017} and experimentally 
	\cite{Jurcevic_2014,Xu_2020}. 
	
	Now that we understand DQPT in terms of symmetry restoration, we introduce the quantifier $\mathcal{R}(t)$ to capture this aspect and detect the DQPT. The quantity $\mathcal{R}(t) $ is defined as:
	\begin{equation}\label{eq:Rate}
		\mathcal{R}(t)=-\frac{1}{N}\ln \left[\prod_{k_n}\left(1-\lvert \langle 
		\psi_{k_n}^G(t)|\sigma_x| 
		\psi_{k_n}^G(t) \rangle \rvert^2\right)\right]
	\end{equation}
	where $ |\psi_{k_n}^G(t) \rangle  $ is the instantaneous ground state of 
	$\vec{d}_{k_n}(t)\cdot\vec{\sigma}$. If we denote the polar angle of the Pauli 
	vector by $\Theta_{k_n}(t)$, then $\left(1-\lvert \langle 
	\psi_{k_n}^G|\sigma_x| \psi_{k_n}^G \rangle \rvert^2\right) = 1 - \sin^2 
	\Theta_{k_n}(t)\cos^2 \Phi_{k_n}(t)$, which reduces to $1- \cos^2 \Phi_{k_n}(t) 
	$ for dynamical critical mode and to 0 for a DQPT to occur. Using the 
	fact that $ |\psi_{k_n}^G(t) \rangle  $ is the ground state of 
	$\vec{d}_{k_n}(t)\cdot\vec{\sigma}$ and 
	$\vec{d}_{k_n}(t)\cdot\vec{d}_{k_n}(t)=1$, standard  algebraic 
	manipulations show that the quantity $1-\lvert \langle 
	\psi_{k_n}^G(t)|\sigma_x| \psi_{k_n}^G(t) \rangle \rvert^2$ is equal to the 
	mode-resolved return probability. This implies that, irrespective of the 
	quench protocol, $\mathcal{R}(t)$ is equivalent to the rate function 
	$r(t)$; more precisely, they differ only by a numerical factor of 2 
	\cite{supp}. This provides a method to derive the rate 
	function based on the principle of symmetry restoration.
	
	It is important to note that $\mathcal{R}(t)$ is designed to capture the 
	dynamical symmetry restoration in time-evolved modes. Its equivalence to 
	the rate function relates the symmetry restoration aspects of 
	mode-dynamics with the statistical understanding of DQPT in terms of 
	dynamical free energy.  This firmly puts across the idea that DQPT occurs 
	at the point of dynamical symmetry restoration in the time-evolved modes.
	
	\noindent \textbf{\emph{$XY$ Model.--}}
	We take the quantum $XY$ model, a prototypical example to study DQPT, 
	which can also be 	experimentally realized in trapped ion systems 
	\cite{Islam_2011} and ultra-cold atoms in optical lattices 
	\cite{Simon_2011}, to further explore the behaviour of DMEs and DQPT. 	
	The Hamiltonian of one-dimensional $XY$ chain of length  $N$ is given by,
	\begin{equation}\label{eq:XY_ham}
		H=-J\sum_{i=1}^N \left[(1+\Delta) S_i^x S_{i+1}^x+(1-\Delta) S_i^y S_{i+1}^y+\mu S_i^z\right],
	\end{equation}
	where the spin operators at $i\,$th site are defined in terms of the Pauli 
	matrices as $S_i^p=\frac{1}{2} \sigma_i^p$ ($p=x,y,z$). Here, $J$ 
	represents longitudinal spin-spin couplings, $\Delta$ governs the 
	anisotropic coupling strength between spins, while $\mu$ denotes the 
	strength of the external transverse magnetic field. We take periodic 
	boundary conditions, \emph{i.e.} 
	$\sigma_{N+1}^p=\sigma_1^p$. The Hamiltonian in Eq.~\ref{eq:XY_ham} can be 
	expressed in terms of the fermionic operators via the Jordan-Wigner 
	transformation, and can be written in momentum space as,
	\begin{equation}\label{eq:XY_momspcace}
		H=\sum _{n=0}^{N-1} \begin{bmatrix}
			c_{k_n}^\dagger& c_{-k_n}
		\end{bmatrix}\begin{bmatrix}
			-\left(\cos k_n+\mu\right) & i\Delta \sin k_n\\-i\Delta \sin k_n & \cos k_n+\mu
		\end{bmatrix}\begin{bmatrix}
			c_{k_n}\\ c_{-k_n}^\dagger
		\end{bmatrix},
	\end{equation}
	where the lattice momenta is quantized as $k_n=\frac{2 \pi }{N} 
	\left(n+\frac{1}{2}\right)$ corresponding to the anti-periodic boundary 
	condition. By comparing, Eq.~\ref{eq:gen_momspace} and 
	Eq.~\ref{eq:XY_momspcace}, we can easily identify 
	$\epsilon_{k_n}=-\left(\mu+\cos k_n\right), \quad \Delta_{k_n}^*=i \Delta 
	\sin k_n$. We consider the quench in both $\mu$ and $\Delta$ for all the 
	calculations and figures unless stated otherwise. We explore the 
	single-parameter quench protocols i.e., quenches in either $\mu$ or 
	$\Delta$ in the Ref.~\cite{supp}.
	\begin{figure}[t]
		\centering
		\includegraphics{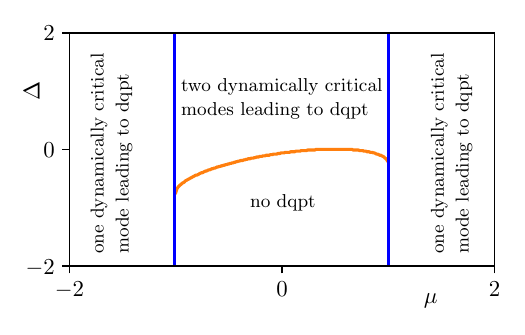}
		\caption{Plot of the post-quench parameters $(\mu, \Delta)$ for which 
			DQPT occurs with either a single or double dynamical critical mode, for 
			fixed pre-quench parameters $(\mu_0,\Delta_0)=(0.5,-1)$.}
		\label{fig:dqptpostquenchpointsalpinf}
	\end{figure}
	A mode becomes dynamically critical at times $t = t_c$ (defined in Eq. 
	\ref{eq:dqpt_cond}). However, for the system to undergo a DQPT, we also 
	need the condition $\cos 2\delta\theta_{k_c}\overset{!}{=} 0$. This 
	condition, in terms of quench parameters, is given by:
	\begin{align}\label{eq:DQPT_condition}
		\begin{aligned}
			\frac{\left(\mu+\cos k_c\right)\left(\mu_0+\cos k_c\right)+\Delta 
				\Delta_0 \sin^2 k_c}{\lambda^{(0)}_{k_c}\lambda_{k_c}} 
			\overset{!}{=} 0.
		\end{aligned}
	\end{align}
	Treating discrete momentum modes as continuous variables in the large 
	system-size limit, we find the momentum modes that satisfy the above 
	equation for the case $\Delta \Delta_0 \neq 1$ as follows:
	\begin{align}\label{eq:cont_k}
		\begin{aligned}
			\cos k_c=-\frac{\left(\mu+\mu_0\right)}{2 \left(1-\Delta 
				\Delta_0\right)}\left[1 \mp 
			\sqrt{1-\Xi\left(\mu_0,\Delta_0,\mu,\Delta\right)}\right],\\
			\text{where}, \Xi\left(\mu_0,\Delta_0,\mu,\Delta\right)=\frac{4 
				\left(1- \Delta \Delta_0\right)\left(\mu \mu_0+\Delta 
				\Delta_0\right)}{\mu+\mu_0}.
		\end{aligned}
	\end{align}
	Depending on the quench protocol, the number of dynamical critical modes 
	which leads to a DQPT (i.e. satisfying Eq. \ref{eq:cont_k}) can be zero, 
	one or two.The number of these modes is plotted in the post-quench 
	$\mu-\Delta$ plane in Fig.~\ref{fig:dqptpostquenchpointsalpinf} for a fixed 
	pre-quench parameters $(\mu_0, \Delta_0) = (0.5, -1)$. 
	
	For the specific case where $\Delta \Delta_0=1$, there exist only a single 
	dynamical critical mode. This occurs when the values of the chemical 
	potential, $\abs{\mu}$ and $\abs{\mu_0}$ are chosen from different sides of 
	critical value 1. This observation implies that quenching across the 
	critical line is necessary for a DQPT to occur when $\Delta \Delta_0=1$. 
	The same feature was observed in Ref.~\cite{Heyl_2013} for the transverse 
	field Ising model, where quench was introduced only in chemical potential. 
	This formed the basis of the argument for DQPT to be connected to ground 
	state quantum phase transition and why for a DQPT to occur critical line 
	$\mu=\pm 1$ needed to be crossed during the quench.
	
\begin{figure}[t]
		\centering
		\includegraphics[width=\linewidth]{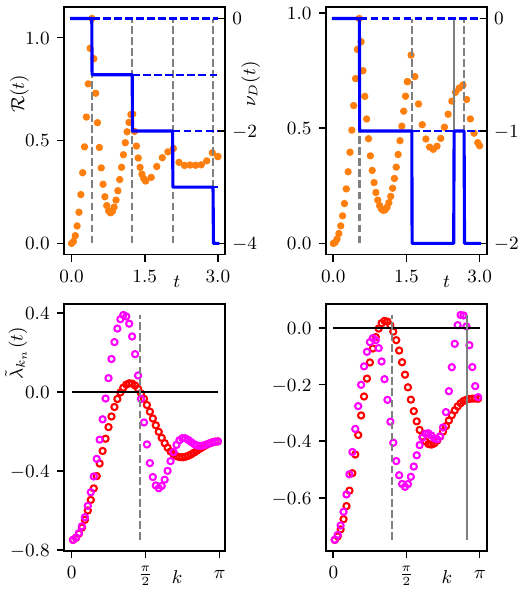}
		\caption{In the left column (top panel), we plot $\mathcal{R}(t)$ 
			(left axis) and DTOP ($\nu_D(t)$) (right axis) as a 
			function of time $t$ for the quench protocol 
			$(\mu_0,\Delta_0)=(0.5,-1)\rightarrow (\mu,\Delta)=(1.5,1)$, where a 
			single mode satisfies the condition $\cos\!\left(2\delta 
			\theta_{k_c}\right)=0$. In the right column, we show the same plot for 
			$(\mu_0,\Delta_0)=(0.5,-1) \rightarrow (\mu,\Delta)=(0.8,1)$, where two 
			modes satisfy the DQPT condition.
			In the left column (bottom panel), for the same quench protocol as in 
			the left column (top panel), we plot the dynamical mode energy 
			$\Tilde{\lambda}_{k_n}(t)$ as a function of momentum $k$ at $t=t_c$ 
			from Eq.~\ref{eq:dqpt_cond}, with $m_2=0$ (red circles) and 
			$m_2=2$ (magenta circles). In the right column, for the same quench 
			protocol as in the right column (top panel), we plot the same at two 
			different $t=t_c$ corresponding to two different $k_c$ values.}
		\label{fig:rfvsttddispvsk}
	\end{figure}

	By also quenching the $\Delta$ parameter, such that pre and post-quench 
	values have different signs,  it's possible to induce a DQPT with two 
	dynamically critical momentum modes even without crossing the $\mu=\pm 1$ 
	critical line.	Allowing the quench in $\Delta $ revealed the situations 
	when DQPT 
	occurs even if corresponding static problem does not show a ground 
	state quantum phase transition.  To illustrate, this connection of single  
	vs double dynamically critical mode with a DQPT in the two different sudden 
	quench scenarios,  i.e. from $(\mu, \Delta) = (0.5, -1)\to (1.5, 1)$ 
	(crossing $\mu=1$ critical line) and  $(0.5, -1)\to (0.8, 1)$ (not crossing 
	the $\mu=1$ critical line), we plot $\mathcal{R}(t)$ and 
	DTOP vs $t$ to observe a DQPT in the top-panel of Fig. 
	\ref{fig:rfvsttddispvsk} \cite{supp}. In the 
	bottom 
	panel, we plot DME for all the $k$-modes at two different critical times 
	($m_2=0, 2$) in bottom-left panel of Fig.~\ref{fig:rfvsttddispvsk} while at 
	two different critical times corresponding to two dynamical critical modes 
	leading to DQPT with $m_2=0$ in the bottom-right panel. The 
	top panels clearly show that 
	$\mathcal{R}(t)$ captures the DQPT in both the sudden 
	quench scenarios. The DTOP exhibits only negative jumps at 
	the critical times for the quench protocol with a single dynamically 
	critical mode leading to DQPT, whereas for quench protocols with two such 
	dynamically critical modes, the DTOP shows both positive and negative jumps 
	at the corresponding two critical times when DQPT occurs. The lower left 
	panel shows that the same mode softens at 
	both critical times for a DQPT to occur when the quench protocol involves 
	crossing the $\mu=1$ critical line. However, there exists other modes where 
	DME vanishes but a DQPT does not occur, and the positions of these modes 
	change at two different critical times with different $m_2$. On the other 
	hand, when we don't cross the $\mu=1$ line during the quench, there 
	are two different modes which become 
	dynamically critical and satisfy Eq. \ref{eq:dqpt_cond}.
	
\begin{figure}[t]
		\centering
		\includegraphics{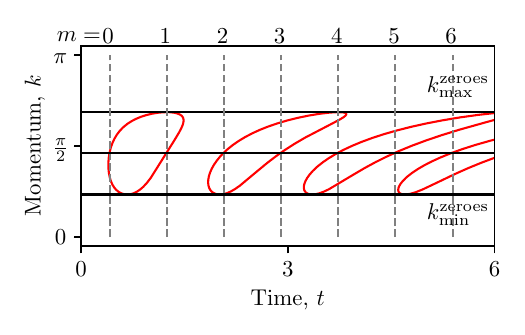}
		\caption{DME zeros in the $k-t$ plane for the same quench protocol as 
			in the left column of fig.~\ref{fig:rfvsttddispvsk}. The three 
			horizontal black lines indicate the momentum modes satisfying 
			$\cos\left(2\delta \theta_{k}\right) = 0$ and $\cos\left(2\delta 
			\theta_{k}\right) = \pm \frac{1}{\sqrt{2}}$. The Black vertical lines 
			denote the times at which $\Phi_{k_c}(t_c)=0$.}
		\label{fig:dme_zero_crossings_3dplot}
	\end{figure}

	Let's analyze another, quench protocol studied in literature 
	\cite{Vajna_2014}, ($\mu_0>1, \Delta=\Delta_0 \to |\mu|<1,\Delta=0$) where 
	despite 
	crossing the critical point, $\mu=1$, no DQPT was observed. 
	For this quench protocol, Eq.~\ref{eq:DQPT_condition} takes 
	the simpler form,
	\begin{equation}
		\left(\mu+\cos k\right)\left(\mu_0+\cos k\right)=0.
	\end{equation}
	Thanks to $\mu_0>1$, the only momentum mode where $\cos 
	2\delta\theta_{k_c}=0$, is given by $k_c=-\cos^{-1} (\mu)$. However, for 
	this mode,  the post-quench mode energy $\lambda_k$ vanishes, causing the critical time to diverge to infinity according to Eq.~\ref{eq:dqpt_cond}. 
	Therefore, for any finite time $t$, DQPTs cannot be observed for this 
	quench protocol. The absence of a DQPT can also be argued alternatively by 
	noting the fact that the Pauli vector for the momentum mode $k=k_c$ stays 
	in its initial position in this quench protocol.  
	However, this does not rule out the vanishing dynamical mode energies for 
	other modes at a 
	finite time, as we need only $d^z_{k_n}$ to vanish. The condition for 
	vanishing $\Tilde{\lambda}_{k_n}(t)$ for this quench 
	protocol is
	\begin{equation}
		\frac{\left(\mu_0+\cos k\right)^2}{\Delta_0^2 \sin^2 (k)}=-\cos 
		\left(2\lambda_k t\right).
	\end{equation}
	The above equation may not be satisfied for all momentum modes if $\mu_0 > 
	\Delta_0$. However, if $\mu_0$ and $\Delta_0$ are comparable, 
	then there may exist critical mode(s) that will satisfy the above equation. 
	For example, if $\mu_0^2 \leq \Delta_0^2$ then $k=\pi/2$ will 
	be the dynamical critical mode with a finite critical time,
	\begin{equation}
		t_c=\frac{1}{2\abs{\mu}}\cos^{-1} 
		\left(-\frac{\mu_0^2}{\Delta_0^2}\right).
	\end{equation}
	Note that this solution is not consistent with Eq.~\ref{eq:dqpt_cond} and, 
	therefore, even though we have vanishing DME at finite time, there won't be 
	a symmetry restoration and consequently, no DQPT. This example clearly 
	demonstrates  the existence of dynamical criticality that does not lead to a DQPT. 
	The dynamical critical mode or zero-modes also have the maximum single mode 
	entanglement; therefore counting these zero-modes with time directly gives 
	entanglement growth in the system. To this end, without loss of generality, 
	we choose a quench protocol in which only one dynamically critical mode 
	occurs which leads to a DQPT and plot DME zeroes in the $k-t$ space in 
	Fig. \ref{fig:dme_zero_crossings_3dplot}. 
	We also mark the DQPT times in Fig. \ref{fig:dme_zero_crossings_3dplot} 
	with dashed lines. The central horizontal solid line correspond to the $k-$ 
	value for which $\cos 2\delta\theta_{k_c}=0$. In $k-t$ space, the DME zeros 
	occur only for $k_\text{min}^\text{zeros} \leq k \leq 
	k_\text{max}^\text{zeros}$. These bounds are quench dependent and can be 
	derived by setting $d^z_{k_n}=0$. For non-critical modes of pre-quench 
	Hamiltonian, using the range of the cosine function, we can show that 
	$|\cos(2\delta\theta_k)|\leq \frac{1}{\sqrt{2}}$.
	The corresponding $k-$values are plotted as two horizontal lines in Fig. 
	\ref{fig:dme_zero_crossings_3dplot}.  
	Interestingly, the areas bounded by the DME zero curve in the $k-t$ space 
	are equal. Beyond the time corresponding to the start of the second DME 
	zero contour, there exists dynamical critical modes for all time and they 
	are 
	even in number. The number of dynamical critical modes, on  average,
	increases linearly with time and so does the momentum space entanglement of 
	the 
	system. In chiral fermions, they contribute to additional non-local 
	source of entanglement in the entanglement Hamiltonian 
	\cite{KlichPRL_2017}.

	\noindent\textbf{\emph{Conclusion.--}} Borrowing the definition of quantum 
	critical modes from ground state quantum phase transition to 
	non-equilibrium domain, we study the dynamical critical modes to 
	understand the dynamical quantum phase transition in the sudden quench 
	dynamics of translationally invariant quadratic fermionic Hamiltonian. We 
	identify the DQPT using the symmetry restoration of $k-$ resolved 
	instantaneous eigenstate of the time evolved pre-quench Hamiltonian. 
	To capture the spin-flip symmetry restoration, we define a quantity 
	$\mathcal{R}(t)$ that exhibits cusp-like singularities at the times when 
	dynamical symmetry restoration or DQPT occurs. We further analytically 
	demonstrate  that $\mathcal{R}(t)$ is proportional to the rate function 
	$r(t)$ for any quench protocol, differing only by an overall factor of 2.
	By establishing this equivalence, we have effectively recovered the rate 
	function through the lens of dynamical symmetry restoration. We have 
	shown that the dynamical critical modes with vanishing DME are necessary but not 
	sufficient for a DQPT. The understanding of when a DQPT and a ground state 
	quantum 
	phase transition occur together and when one occurs without the other is 
	very transparent in terms of dynamical critical modes. Since the 
	dynamical critical modes also have the maximum single mode entanglement, 
	the $k-$ mode entanglement entropy of the system will be bounded from below 
	by the number of these modes times $\ln 2$. As the number of such modes, on 
	average, increases linearly with time, so does the entanglement entropy.

\newpage

\newcommand{\snum}{S}

\renewcommand{\theequation}{\snum.\arabic{equation}}
\renewcommand{\thefigure}{\snum.\arabic{figure}}

\setcounter{equation}{0}
\setcounter{figure}{0}

\newpage 
\section*{Supplementary material: Dynamical quantum phase transitions through 
the lens of mode dynamics}
\label{appendix1}

	\section{Dynamical symmetry restoration and rate function 
	divergence}\label{sec:rf}
	
	The rate function has been the widely used quantity to detect dynamical 
	quantum phase transition (DQPT) 
	\cite{Heyl_2013,Vajna_2014,Heyl_2014,Heyl_2015,Halimeh_2017,Homrighausen_2017,Zauner_Stauber_2017,unkovi_2018,Lang_2018}.
	 For a one-dimensional system of size $N$, it is defined as 
	\cite{Heyl_2018,heyl_2019,Heyl_2013}
	\begin{equation}
		r(t) = -\lim_{N \to \infty} \frac{1}{N} \ln |G(t)|^2,
	\end{equation}
	where $G(t)$ denotes the Loschmidt (or return) amplitude at time $t$, given 
	by
	\begin{equation}
		G(t) = \langle \psi(0) | e^{-i H t} | \psi(0) \rangle.
	\end{equation}
	Here, $|\psi(0)\rangle$ represents the initial state, and $H$ is the 
	post-quench Hamiltonian. The expression for the rate function is already 
	known from Ref.~\cite{Heyl_2013} and is given by 
	\begin{equation} \label{eq:rf}
		r(t) = -\frac{1}{2\pi}\int_0^\pi \ln \left[1 + 4p_k(p_k - 1)\sin^2 
		\left(2\lambda_k t\right)\right] dk,
	\end{equation} 
	where $p_k = \sin^2 \left(\delta \theta_k\right)$.

	As discussed in the main text, starting from a symmetry-broken state, DQPT 
	occurs for a given momentum mode and time when the instantaneous ground 
	state of $\vec{d}_{k_n }(t)\cdot\vec{\sigma}$ becomes symmetric under the 
	transformation that flips the pseudospin eigenvectors, i.e., 
	$|\uparrow\rangle \leftrightarrow |\downarrow\rangle$. Under such 
	circumstances, the polar angle $\Theta_{k_n}(t)$ takes the value $\pi/2$, 
	while the azimuthal angle $\Phi_{k_c}(t_c)$ vanishes. Consequently, 
	$d^y_{k_c}(t_c) = 0$, $d^z_{k_c}(t_c) = 0$, and $d^x_{k_c}(t_c) \neq 0$. 
	Following Eq.~(9) of the main text, $d^y_{k_c}(t_c)$ can vanish under three 
	possible conditions:  
	(i) $\sin(2\delta \theta_{k_n})=0$, which, however, makes 
	$d^x_{k_c}(t_c)=0$;  
	(ii) $\cos(2\lambda_{k_n} t)=1$, which gives $d^z_{k_c}(t_c)\neq 0$ 
	irrespective of the quench protocol;  
	(iii) $\cos(2\delta \theta_{k_n})=0$, which is the only valid condition. To 
	further ensure $d^z_{k_c}(t_c)=0$, we require $\cos(2\lambda_{k_n} t)=0$. 
	If we put these two conditions in Eq.~\ref{eq:rf}, then the rate function 
	diverges. Therefore, the momentum mode and time at which the instantaneous 
	ground state of $\vec{d}_{k_n }(t)\cdot\vec{\sigma}$ becomes symmetric must 
	correspond to a divergence in the rate function.
	\begin{figure}
		\centering
		\includegraphics[width=0.7\linewidth]{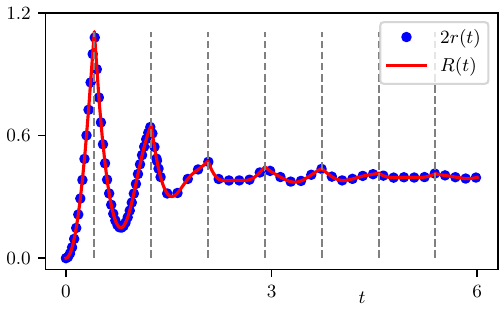}
		\caption{Plot of $\mathcal{R}(t)$ (red solid line) and $2r(t)$ (blue 
		circles) as functions of time for the quench protocol 
		$(\mu_0,\Delta_0)=(0.5,-1) \rightarrow (\mu,\Delta)=(1.5,1)$, where 
		only a single mode $k=k_c$ exists such that $\cos\left(2\delta 
		\theta_{k_c}\right)=0$. The vertical dotted lines correspond to the 
		critical times $t=t_c$ given in Eq.~(11) of main text, with $m_2 \in 
		[0,6]$.}
		\label{fig:rf_sxexp_vs_t_new}
	\end{figure}
	\newpage
	\section{Equivalence between $R(t)$ and $r(t)$}
	\noindent In this section, we explore the connection between the quantity 
	$\mathcal{R}(t)$, defined to identify the symmetry restoration points and 
	the rate function $r(t)$. Let $|\psi\rangle$ be an eigenvector of the 
	operator $\overrightarrow { d_{k_n} }(t) \cdot \overrightarrow { \sigma } 
	$, such that
	\begin{equation}
		\overrightarrow { d_{k_n} }(t) \cdot \overrightarrow { \sigma }  | 
		\psi  \rangle = \lambda | \psi  \rangle.
	\end{equation}
	Applying $\overrightarrow { d_{k_n} }(t) \cdot \overrightarrow { \sigma } $ 
	once more on both sides and utilizing,
	\begin{equation}\label{eq:unit_condition}
		(d_{k_n}^x(t))^2+(d_{k_n}^y(t))^2+(d_{k_n}^z(t))^2=1,
	\end{equation} 
	we get $\lambda^2=1$, implying $\lambda = \pm 1$. We denote $| 
	\psi_{k_n}^G(t) \rangle$ as the eigenvector corresponding to the lower 
	eigenvalue $\lambda=-1$. We now show that
	\begin{equation}
		\langle \psi_{k_n}^G(t)| \sigma_{ x }|\psi_{k_n}^G(t)\rangle = 
		-d_{k_n}^x(t).
	\end{equation} 
	Using the anti-commutation relation of the Pauli matrices, we can write
	\begin{equation}
		\{ \overrightarrow { d_{k_n} }(t) \cdot \overrightarrow { \sigma }, 
		\sigma_x\} = 2d_{k_n}^x(t).
	\end{equation}
	After taking the expectation value of the above identity with respect to 
	the state $|\psi_{k_n}^G(t)\rangle$, we obtain
	\begin{equation}
\begin{aligned}
		\langle \psi_{k_n}^G(t)| 	(\overrightarrow { d_{k_n} }(t) \cdot 
		\overrightarrow { \sigma })  \sigma_x + \sigma_x  (\overrightarrow { 
		d_{k_n} }(t) \cdot \overrightarrow { \sigma }) |\psi_{k_n}^G(t)\rangle  
		=\\
		 2d_{k_n}^x(t) \langle \psi_{k_n}^G(t)|\psi_{k_n}^G(t)\rangle.
\end{aligned}
	\end{equation}
	Since $(\vec{d}_{k_n}(t)\cdot \vec{\sigma})|\psi_{k_n}^G(t)\rangle = 
	-|\psi_{k_n}^G(t)\rangle$, the above relation can be simplified as
	\begin{equation}\label{eq:simpl}
		\langle \psi_{k_n}^G(t)| \sigma_x |\psi_{k_n}^G(t)\rangle = - 
		d_{k_n}^x(t).
	\end{equation}
	As a consequence of Eq.~\ref{eq:simpl}, we find
	\begin{equation}
		\begin{aligned}
			\left(1-\lvert \langle 
			\psi_{k_n}^G(t)|\sigma_x| \psi_{k_n}^G(t) \rangle 
			\rvert^2\right)&=1-(d_{k_n}^x(t))^2\\ &= 1-\sin^2 \left(2 \delta 
			\theta_{k_n}\right) \sin^2 \left(2\lambda_{k_n} 
			t\right)\\
			&  = 1- 4 \sin^2( \delta \theta_{k_n})\cos^2( \delta 
			\theta_{k_n})\sin^2 \left(2\lambda_{k_n} 
			t\right)\\
			& =  1+4p_{k_n}(p_{k_n}-1) \sin^2 \left(2 \lambda_{k_n}t\right)
		\end{aligned}
	\end{equation}
	As can be seen from Eq.~\ref{eq:rf}, the right-hand side of the above 
	expression corresponds exactly to the mode-resolved return 
	probability. Therefore, we analytically establish that the quantities 
	$\mathcal{R}(t)$ and $r(t)$ are equivalent. They differ only by an overall 
	factor of 2, since in calculating $\mathcal{R}(t)$ we consider all the 
	momentum modes in the interval $[0,2\pi]$ , while in the rate function, the 
	momentum modes are in the interval $[0,\pi]$. Numerically, the equivalence 
	between $2r(t)$ and $\mathcal{R}(t)$ is confirmed in 
	Fig.~\ref{fig:rf_sxexp_vs_t_new}, where an exact agreement between these 
	two is found.

	\section{Dynamical topological order parameter (DTOP)}
	
	In this section, we recall the definition of dynamical topological order 
	parameter (DTOP) and investigate the effect of spin-flipping symmetry 
	restoration of the instantaneous eigenstates of dynamically evolved 
	pre-quench Hamiltonian on this. 
	The DTOP has been widely used in several works to characterize DQPT  
	\cite{Budich_2016,dutta_2017,Utso_2017,Yang_2019} and has been measured 
	experimentally in quantum walks \cite{Xu_2020}. 
	
	To define the DTOP, we require the Pancharatnam Geometric Phase (PGP), 
	originally introduced for light beams with non-orthogonal polarization 
	\cite{Pancharatnam_1956}. Importantly, The PGP can be defined smoothly only 
	when the initial and the time-evolved state remain non-orthogonal. The PGP 
	is defined by subtracting the dynamical phase from the total phase of the 
	mode-resolved Loschmidt amplitude. Specifically, 
	\begin{equation}
		\phi_k^G(t)=\phi_k(t)-\phi_k^{\rm dyn}(t),
	\end{equation}
	where $\phi_k^{\rm dyn}(t)=-\int_0 ^t \langle \psi(s)|H|\psi(s)\rangle ds$ 
	is the dynamical phase, and the total phase $\phi_k(t)$ is obtained by 
	writing the mode-resolved Loschmidt amplitude ($G_k$) in polar form as 
	$G_k(t)=g_k(t) e^{i \phi_k(t)}$. The full Loschmidt amplitude is given by 
	$G(t)= \prod_k G_k(t)$. The DTOP is then defined from the PGP as 
	\cite{Budich_2016}
	\begin{equation}\label{eq:DTOP}
		\nu_D=\frac{1}{2 \pi } \int_0^\pi\frac{\partial\phi_k^G(t)}{\partial k} 
		dk.
	\end{equation}
	As discussed in Sec.~\ref{sec:rf}, dynamical symmetry restoration of the 
	instantaneous ground state of $\vec{d}_{k_n }(t)\cdot\vec{\sigma}$ occurs 
	when $d^y_{k_c}(t_c) = 0$, $d^z_{k_c}(t_c) = 0$. Following 
	Eq.~\ref{eq:unit_condition}, this naturally implies $(d^x_{k_c}(t_c))^2 = 
	1$. These conditions are satisfied only when $\cos(2 \delta \theta_k)=0$ 
	and $\cos (2 \lambda_k t)=0$. When these two conditions are simultaneously 
	satisfied, from Eq.~\ref{eq:rf}, it can be easily seen that $g_k(t)$ of the 
	mode resolved Loschmidt amplitude vanishes. Under such circumstances, the 
	PGP is not well defined and therefore the DTOP in Eq.~\ref{eq:DTOP} makes 
	an integer jump. This explains the integer jumps obtained at DQPT times in 
	Fig.~2 for both quench protocols in the main text.

	\section{Mode dynamics and
		DQPT for Single Quench Protocols}
	
	In this section, we complement our results for quenches in both $\mu$ and 
	$\Delta$ by examining the single-parameter quench protocols, i.e., quenches 
	in either $\mu$ or $\Delta$ while keeping the other parameter fixed. For 
	the double quench protocols, we showed that although many momentum modes 
	may correspond to vanishing energy, only a subset of these modes are 
	special in the sense that their eigenvectors respect the spin flipping 
	symmetry. The existence of these symmetry-restoring modes leads to the 
	divergences in the rate function equivalent quantity $\mathcal{R}(t)$ and 
	produces unit jumps in the DTOP. 
	
\begin{figure}
		\centering
		\includegraphics[width=0.7\linewidth]{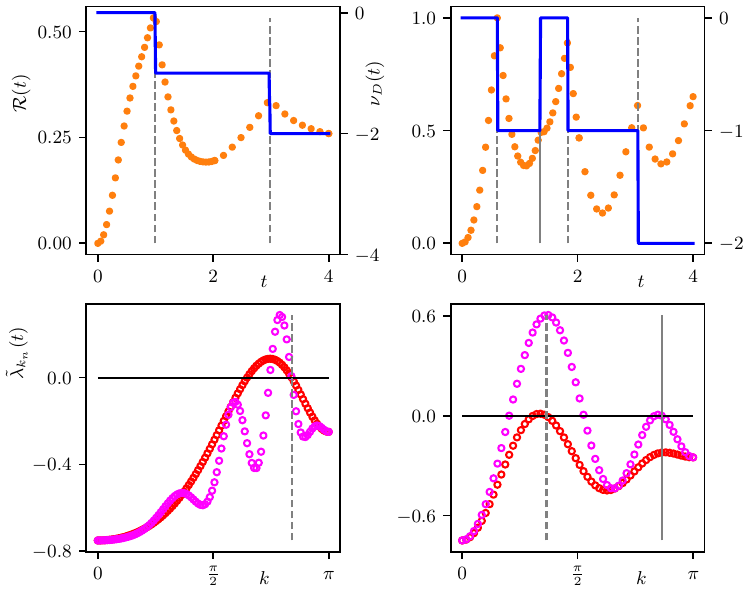}
		\caption{(Top left), plot of $\mathcal{R}(t)$ (left axis) and the DTOP 
		(right axis) as a function of time for the single-quench protocol 
		$\mu_0=0.5,\mu=1.5$ and $\Delta=\Delta_0=1$.  In the right column, we 
		plot the same for the quench protocol where $\mu=\mu_0=0.5$ and 
		$\Delta_0=-1, \Delta=1$. In this case, two momentum modes satisfy the 
		dynamical symmetry restoration condition, and the associated critical 
		times are indicated by solid and dashed vertical lines. For the same 
		quench protocol as in the top-left panel, in the bottom-left panel, we 
		plot the DME as a function of momentum modes at the critical times with 
		$m_2=0$ (red circles) and $m_2=2$ (magenta circles), corresponding to 
		the single critical momentum mode that satisfies $\cos(2 \delta 
		\theta_{k_c})=0$. This mode is indicated by the vertical dotted line. 
		In the right column, we plot the same for a fixed chemical potential 
		$\mu_0=\mu=0.5$ and performing the quench in $\Delta$ from 
		$\Delta_0=-1$ to $\Delta=1$ at two critical times with $m_2=0$ 
		associated with the two critical momentum modes.}
		\label{fig:rfvsttddispvsk1sq}
	\end{figure}

	To test the robustness of these observations for single-parameter quenches, 
	we choose two quench protocols. In the first protocol, we quench the 
	chemical potential from $\mu_0=0.5$ to $\mu=1.5$, while $\Delta$ is fixed 
	at $\Delta=\Delta_0=1$. In this case, the critical momentum mode that 
	satisfies $\cos(2\delta \theta_k)=0$ is found to be 
	\begin{equation}\label{eq:kc}
		k_c=\frac{-(1+\mu\mu_0)}{\mu+\mu_0}.
	\end{equation} 
	At the corresponding critical time $t=t_c$, the quantity $\mathcal{R}(t)$ 
	diverges and the DTOP undergoes a negative unit jump, as shown in the 
	top-left panel of Fig.~\ref{fig:rfvsttddispvsk1sq}. The mode $k=k_c$ is 
	represented by a vertical dotted line in the bottom-left panel of the same 
	figure. Although there exist other momentum modes with vanishing DME at 
	both the critical times associated with $m_2=0$ and $m_2=2$, these modes do 
	not induce symmetry restoration.

	In the second quench protocol, we fix the chemical potential at 
	$\mu=\mu_0=0.5$ and quench $\Delta$ from $\Delta_0=-1$ to $\Delta=1$. In 
	this case, there are two critical modes that satisfy the condition 
	$\cos(2\delta \theta_k)=0$. Their values, obtained from Eq.~16 of the main 
	text, are marked by dotted and solid vertical lines in the bottom-right 
	panel of Fig.~\ref{fig:rfvsttddispvsk1sq}. The associated critical times 
	are shown by corresponding vertical lines in the top-right panel. At both 
	critical times, $\mathcal{R}(t)$ diverges. The DTOP makes a unit positive 
	jump at critical times associated with one of the symmetry-restoring modes 
	and a negative jump at the critical times associated with the other. 
	Similar to the previous case, there exist other momentum modes that 
	correspond to vanishing DME but their eigenvectors do not respect the 
	spin-flipping symmetry.

	\bibliography{DQPT_refs.bib}

\end{document}